\documentclass[conference]{IEEEtran}
\IEEEoverridecommandlockouts
\usepackage{cite}
\usepackage{amsmath,amssymb,amsfonts}
\usepackage{algorithmic}
\usepackage{graphicx}
\usepackage{textcomp}
\usepackage{xcolor}
\usepackage{todonotes}
\usepackage{hyperref}
\usepackage{multirow}
\def\BibTeX{{\rm B\kern-.05em{\sc i\kern-.025em b}\kern-.08em
    T\kern-.1667em\lower.7ex\hbox{E}\kern-.125emX}}
\begin{document}

\title{Techniques and Applications for Crawling, Ingesting and Analyzing Blockchain Data\\
}

\author{
  \IEEEauthorblockN{
    Evan Brinckman\IEEEauthorrefmark{1},
    Andrey Kuehlkamp\IEEEauthorrefmark{1}\IEEEauthorrefmark{3},
    Jarek Nabrzyski\IEEEauthorrefmark{1},
    Ian J. Taylor\IEEEauthorrefmark{1}\IEEEauthorrefmark{2}
  }
  \IEEEauthorblockA{
     \IEEEauthorrefmark{1}Center for Research Computing, University of Notre Dame, Notre Dame, IN, USA \\
     \IEEEauthorrefmark{2}School of Computer Science \& Informatics, Cardiff University, Cardiff, UK \\
    \IEEEauthorrefmark{3}\emph{Corresponding Author}
  }
}

\maketitle

\begin{abstract}
As the public Ethereum network surpasses half a billion transactions and enterprise Blockchain systems becoming highly capable of meeting the demands of global deployments, production Blockchain applications are fast becoming commonplace across a diverse range of business and scientific verticals. In this paper, we reflect on work we have been conducting recently surrounding the ingestion, retrieval and analysis of Blockchain data.  We describe the scaling and semantic challenges when extracting Blockchain data in a way that preserves the original metadata of each transaction by cross referencing the Smart Contract interface with the on-chain data.  We then discuss a scientific use case in the area of Scientific workflows by describing how we can harvest data from tasks and dependencies in a generic way.  We then discuss how crawled public blockchain data can be analyzed using two unsupervised machine learning algorithms, which are designed to identify outlier accounts or smart contracts in the system.  We compare and contrast the two machine learning methods and cross correlate with public Websites to illustrate the effectiveness such approaches.   
\end{abstract}


\section{Introduction}

At the time of writing, the public Ethereum network has been running for around 1450 days since July 2015, and according to \cite{etherscan-transactions} it has processed almost 506 million transactions to date.  The average number of daily transactions is 350,000 since its inception, and recently in June 2019, \cite{cointelegraph} noted that the number of transactions exceeded 1 million again for the first time since the cryptocurrency boom ended in May 2018. For enterprise deployments, companies tend to deploy their private Blockchain ledgers and organizations, such as Enterprise Ethereum Alliance \cite{entethallianc}, have 250 member companies that are actively pursuing private Blockchain deployments. Other enterprise open source Blockchain platforms that are based on Ethereum, such as Quorum, are specifically designed for Enterprise use, combining  the innovation of the public Ethereum community with enhancements to support enterprise needs.  Quorum's vision is to enable solutions that allow companies to securely share their Blockchain business processes and data with other partners and consequently include node based authorization to define which companies have access to what. To this end, Quorum has partnered with Microsoft to provide Quorum as an elastic Azure Blockchain Service \cite{azure-blockchain} earlier this year.  Such elastic private deployments are leading to multiple innovative business applications, which, in turn, will lead to a plethora of Blockchain data that contains a valuable transaction history of business, where insights can be drawn, and fed into new business models that can continue to drive innovation.  

Similarly, in research, there is a broad range of interest in Blockchains, from the low level mechanisms including novel consensus algorithms, to applied research in a multitude of domains where Blockchain has become a research topic in understanding the qualitative and quantitative benefits that impact such sectors.   

In this paper, we describe three studies we've been conducting recently surrounding the ingestion, retrieval and analysis of Blockchain data.  The research began with an investigation into how one might extract data that has been recorded on an Ethereum Blockchain, in order to better understand Ethereum transactions, the encoding used and how the original metadata of transactions might be retrieved in full.  The state of Ethereum Blockchain changes continually through activities such as transactions, contracts, and mining. Ethereum data structure  is very complex, and it
uses the trie data structure to manage the relationship between confirmed transactions 
balances and smart contracts' data. 
We discuss this in detail in 
Section \ref{sec:crawling}.

The second use case we consider is the ingesting of data into the Blockchain in the context of executable science experiments, using computational workflows. Workflows provide a structured methodology for describing complex functional execution, data, and logic pipelines. They enable the reproducibility of experiments and results by exposing underlying scientific processes in a detailed and reproducible fashion. Scientific and business processes can be modeled as a set of self-contained tasks, which can be independently developed, validated, refined and composed for different configurations~\cite{Deelman2009} and by specifying the control of data dependencies and logic for execution, workflows are capable of modeling the entire scientific process. Workflow uptake has increased over the past several years and systems have become very sophisticated and are designed to address a wide range of distributed computing and data dependency needs within a vast multitude of application domains. 

Workflows are a good use case for Blockchain ingestion because they are generic platforms that support multiple scientific research studies. Once integrated, insights can be recorded on the Blockchain, then later crawled and analyzed, to verify and validate the processes and cyberinfrastructure used to conduct such research.  To this end, we integrated a Blockchain tracking component into the Pegasus workflow system \cite{deelman2005}.  Pegasus is a DAG-based workflow system that interacts with multiple data management and computational infrastructures and has been used in multiple contexts, including the recent LIGO gravitational wave discovery.  This is described in Section \ref{workflows}.  
    
The third use case we consider is for the analysis of Blockchain data.  In this use case, we used machine learning to extract  potentially suspicious accounts from the public Ethereum Blockchain. To this end,  we used two machine learning algorithms to attempt to discern between normal and nefarious Ethereum accounts  by learning to detect outliers by using clustering on their transactional history. In this study, we used support vector machines (SVM) and K-means clustering to detect such outliers, which may be scam accounts or suspicious, for some other reason.  Our results, presented in Section \ref{sec:anomalies}, indicate a correlation between the outliers we extract and those accounts flagged as suspicious by Etherscan. 

The rest of this paper is organized as follows: Section \ref{sec:related} presents an overview of similar works; our investigations are described in Sections \ref{sec:crawling}, \ref{workflows}, and \ref{sec:anomalies}; finally, we present our conclusions in Section \ref{sec:conclusions}.

\section{Related Work}
\label{sec:related}

Extracting information from the blockchain is not a new idea: efforts started as soon as people realized the difficulties imposed by the nature of the blockchain's structure. Several of those early tools are not available anymore. Perhaps one of the first endeavors was a library of forensic tools for the Bitcoin network called \emph{BitIodine} \cite{Spagnuolo2014}. The tools were capable of parsing the blockchain to collect and visualize relevant information, having also the ability to perform address clustering, classification and labeling. After extracting information from the blockchain, it is stored in a relational database. 
The authors analyzed known cases involved with the black market web site \emph{Silk Road} and victims of the \emph{CryptoLocker} ransomware, where they identified specific connections between the Silk Road and its founder, using only information publicly available on the blockchain. 

Trying to expand and generalize on the idea of extracting useful information from the blockchain, \cite{Kalodner2017} presented \emph{BlockSci}, a platform for blockchain analysis that supports several different blockchains. They use an in-memory database, and claim speedups from 15 to 600x in relation to other approaches. However, it does not provide support to Smart Contract platforms like Ethereum.
In a subsequent similar effort, \cite{Bartoletti2017} proposed a framework to support data analytics on Bitcoin and Ethereum, reorganizing blockchain data in SQL or NoSQL databases. They try to address the need for combining blockchain data with external information such as user, market and crime-related data. 

Correspondingly to other efforts focused on the Bitcoin network, \cite{Bragagnolo2018} proposed \emph{Ethereum Query Language} (EQL), to query the Ethereum blockchain directly using a SQL-like language. The authors pose the problem mentioning the difficulty of access to the blockchain data through any other means than sequential access or indexed hash.  They implement indexing through Binary Search Trees, however, this brings high storage requirements. Despite being conceived for the Ethereum network, the authors acknowledge EQL has limitations to deal with contracts.

In a work that falls in a slightly different category, \emph{Blockbench} \cite{Dinh2017} presents a framework for comparison between private blockchains in terms of data processing capabilities. However it does not approach data extraction directly, it highlights the limitations of blockchains as a direct source for data analytics.  
One of the main conclusions is that these blockchains are not well suited for large scale data processing.  

Several works propose the ingestion of blockchain transactions to a graph-based database, as a means to facilitate data analytics. In one of them, \cite{Ron2013} built a graph of all transactions and addresses in the Bitcoin network up until May 2013. 
They tried to gain an understanding of user profiles and their balances, how they keep, move or exchange their Bitcoin assets. They found that while an enormous number of transactions exchange only fractional amounts, hundreds of transactions transferred more than 50,000 Bitcoin. 

Also using a graph constructed from the blockchain transactions, \cite{Fleder2015} studied the anonymity in the Bitcoin network. Among other noticeable activities, applying the PageRank algorithm classified an account that is associated to the FBI (and to which funds seized from the famous Silk Road were transferred) as highly important.
Similarly, \cite{Chan2017} were able to track assets illegally obtained from the Gatecoin hack being redirected to accounts associated with known cryptocurrency exchanges, and suggests that a graph interpretation of blockchain data may be the more appropriate for this type of application. 

In another approach, \cite{Bistarelli2017} proposed \emph{BlockChainVis}, a tool for visual analysis of Bitcoin transaction flows to identify illegal activity, ingesting the entire Bitcoin transaction history into its database. A flagged set of nearly 16,000 transactions turned out to correspond to exchange of values between only two addresses, however these could not be confirmed to be related to illegal activity.
Also trying to visually identify anomalous patterns in the transactions, as well as empirically verify economic indicators through metrics like inflation and velocity of circulation, \cite{McGinn2018} ingested roughly 8 years of Bitcoin transaction data into a graph database, using a cluster of 400 cores. The authors estimate a total size of the resulting graph at about 600GB. 

Looking for local patterns that can be correlated to important events, like asset price fluctuations, \cite{Akcora2018} extend the concept of network motifs and graphlets to blockchain graphs, to what they call \emph{chainlets}. The authors tested the statistical Granger causality between changes in chainlets and their effect on Bitcoin price, concluding that certain types of chainlets can be used to predict Bitcoin prices.

In a noticeable trend also among the works that focus on extracting or manipulating data from the blockchain, a number of works set their aim specifically in fraud detection. 
\cite{Moser2013} investigate the influence of anonymization services in the Bitcoin network, as many of them can be used in money laundering activities. Their results demonstrate these services add an effective layer of difficulty for most crime investigators. 

On a different direction inside this area, \cite{Toyoda2017} tried to identify financial frauds known as \emph{Ponzi schemes} in Bitcoin. The authors created a set of features extracted from transactions and fed them into Random Forest and XGBoost to obtain a classification result that indicates potential fraud activity. Using this approach, they report a true positive rate of about 83\% with a false positive rate below 4.4\%.

Focusing on the Ethereum network, \cite{Bartoletti2017}  present a methodology to identify and collect information on Smart Contracts that implement Ponzi schemes and investigate their properties. The authors were able to identify an alarming number of Smart Contracts involved in this type of fraud, but fortunately their impact was still small. More recently, \cite{Bartoletti2018} turned to the Bitcoin network to detect activities related to such schemes. 
The authors manually collected a dataset composed by the transaction graphs associated to Ponzi-related addresses, along with a sample of graphs unrelated to such activities.
They used supervised machine learning algorithms to perform detection of Ponzi-related activity. The best model was a Random Forest classifier, with an accuracy of nearly 99\%, however, it is not clear if they made use of cross-validation to prevent overfitting the training set.

Despite the growing number of works in this area and the massive volume of data available in current blockchains, there is a lack of proper machine learning datasets constructed with well defined objectives and shared among the research community. Most of the works create their own dataset out of a  subset of the blockchain. In each of them, the authors simply seem to use criteria they deem appropriate.
Another point that gives rise to further investigation is the fact that most of past works either: a) focus on the Bitcoin network, which does not offer built-in support for Smart Contracts; or b) put aside Ethereum's comprehensive Smart Contract mechanism as a source for data to be extracted and analyzed.

\section{Lessons learned in crawling blockchain data}
\label{sec:crawling}


As the name suggests, blockchain is a sequential data structure by nature. A blockchain like Ethereum consists of collections of transactions, packed into blocks, linked together sequentially and in an immutable manner. The integrity of data is ensured by the use of cryptographic hash functions: any change in the input data will cause the hash function to generate a different value, indicating the data was modified.
This ensures the immutability of the sequence of blocks, along with all their contents.


Additionally, the use of Merkle tree \cite{Wood2018} structures can allow extremely fast integrity verification of the entire blockchain. A Merkle tree is a hierarchical structure where every element is linked to its children by their hashes, down to the leaves, which correspond to the hashes of each block (or transaction). This allows peers in the network to detect tampering in the data almost instantaneously: it is possible to verify if two copies of the entire blockchain are identical just by comparing the root hash of their Merkle trees, regardless of the size of the tree. 

These characteristics that the blockchain a data storage that is at the same time immutable and distributed  because its verification mechanisms eliminate the need for peers to trust each other. However, this architecture also shapes the blockchain as a sequential structure and limit its flexibility. The Ethereum node API, for instance, only allows retrieval of blocks by their sequential number or by their hash. Suppose a block that was written in a specific date has to be retrieved: unless the block hash is known beforehand, the only alternative is to search blocks sequentially (using their number), and verifying their dates, until the desired date is found. 

Search is not the only operation that is hindered by the blockchain structure. Aggregation and join queries are trivial in traditional database management systems but are prohibitive in the blockchain because of the complexity of their implementation and computational cost. This is the reason why virtually all of the previous works in this area perform some type of data extraction and transformation as a preprocessing step for the analysis.

\subsection{Smart contract information}

Smart Contracts are, in a simple way, distributed programs that are stored in the blockchain and essentially provide a data interface to what is stored and enable further logic to be embedded.  The data interface in Ethereum uses functions to define a data transaction payload, along with typed parameters that define what is written and in what format. All the network nodes have access to the same executable code, and Ethereum architecture ensures that the result of that program will be the same for any node that executes it. If the results differ, consensus between the nodes cannot be reached, and one of them will have to be discarded. This also means that anyone can verify the output of a Smart Contract, by executing it with the same inputs. 

Ethereum implements Smart Contracts through its execution model, known as Ethereum Virtual Machine (EVM). In the EVM, Smart Contracts are a special type of account, and they have access to two types of persistent memory storage. The first type is the account storage, a virtually infinite storage that is granted to each account. However, storing data in the account storage is very expensive. The second type is through event logs: during execution, contracts can trigger events that record data in a log. Also, the invoking transaction stores the data that is passed as arguments to the invoked contract.

Extracting information stored in a contract's account storage is possible, but not trivial, especially for non-basic data types. The structure of the storage is defined in the source code of the smart contract, which is not always available. 
The information that is passed in contract invocations is also encoded, and to have access to it, it is required to have either the contract Application Binary Interface (ABI) or the source code itself.

Most blockchain tools are limited to extracting and storing binary transaction information unmodified, but in the case of interaction with smart contracts, this also limits the ability to gain understanding of the data semantics. The binary payload of these transactions contains encoded information about which contract methods are invoked, and the arguments that are passed in these calls. Therefore, it is eminently important for a data extraction tool to be able to semantically reconstruct the information from these interactions whenever the source code or the ABI of the contract is available. 

\subsection{Data ingestion}

Following the trend described in Section \ref{sec:related}, and with the intention to create a more flexible data organization, we have developed a blockchain extraction tool for the Ethereum network that is capable of crawling the blocks and ingesting its contents into a relational database. However, the scale of a public blockchain like Ethereum is a problem by itself: Currently, the Ethereum network encompasses more than 70 million distinct addresses \cite{etherscan-accounts}, more than 500 million transactions in over 8 million blocks \cite{etherscan-transactions}, and a strong growth tendency. 
We set up a fully synchronized local Ethereum node, and to improve efficiency, we built our tool to run multiple threads, potentially multiplying the ingestion rate by the number of simultaneous threads. On the other hand, running a multi-threaded crawler creates additional strain on the Ethereum node and the database server, to the point where I/O started to become a bottleneck.

Efficiently storing and manipulating the entire public Ethereum history in consumer-grade computers is quickly becoming harder. Last year Google announced the public availability of the entire Ethereum history (along with other blockchains) on their cloud computing platform \cite{google-ethereum}. This image is incrementally updated daily. Their platform allows querying the blockchain data on their BigData infrastructure using traditional SQL language. 
Querying Ethereum data from Google BigQuery platform is highly convenient from the data analytics standpoint: complex queries are processed with great performance in a distributed environment, while the costs are fairly low, following the ``pay-as-you-go'' business model. There are however some situations where BigQuery public dataset might not be adequate, including such examples as 1) private blockchains, and 2) exploration/interaction with Smart Contracts. In the first case, it is necessary to extract and transform the blockchain data to be able to perform analysis on it, for the reasons discussed in the beginning of this Section.

\section{Data Ingestion in Science: Providing Audit Trails for Scientific Workflows}
\label{workflows}

Workflows provide a generic means of specifying a set of jobs, along with control or data dependencies, which can be distributed across a large number of distributed resources.  Providing the capability of recording a non repudiable trail of workflows therefore,  can give a number of insights into the performance of specific workflows on different cyber-infrastructures and can provide a useful tool for researchers and consortium to track scientific experiments that have been run, along with the datasets that have been used in each case. For example, the makespan of multiple workflow instances can be analyzed to better understand empirically which distributed architecture suits that computational pipeline better.  Alternatively, one might use such information for validation and verification of the data: recording the hashcodes of the data used in each workflow, and verify the correct datasets were used for a set of experiments.  Further, if a dataset contained an anomaly, one could search and retrieve all workflows that processed that dataset in order to quickly understand which experiments need to be re-run to rectify the results. 

To this end and to demonstrate the feasibility of using Ethereum blockchain to store workflow execution events, we have developed a proof of concept tracking system, using an open-source and well known workflow management system, Pegasus\footnote{\url{https://github.com/pegasus-isi/pegasus}}.  We created an add-on for Pegasus that was configured to run inside a container, along with a local HTCondor\footnote{\url{https://research.cs.wisc.edu/htcondor/}} pool as the execution environment. Test workflows that are bundled with the software were thenrun through Pegasus, while monitoring the transactions being written to the blockchain. The objective was to verify that every stage of the workflow processing is properly logged onto the blockchain.

\subsection{Implementation}

We have modified Pegasus, adding to its logging subsystem the ability to write flow execution events into a private Ethereum network. Using a middleware called \emph{Aladdin}, we enabled Pegasus to send log events to a Smart Contract written with the intent of keeping a record of such events.
Aladdin is an Application Programming Interface (API) developed by the Center for Research Computing to provide interaction with Smart Contracts on the Ethereum network over HTTP.

\vspace{-2mm}
\begin{figure}[hbt]
    \centering
    \includegraphics[width=1\linewidth]{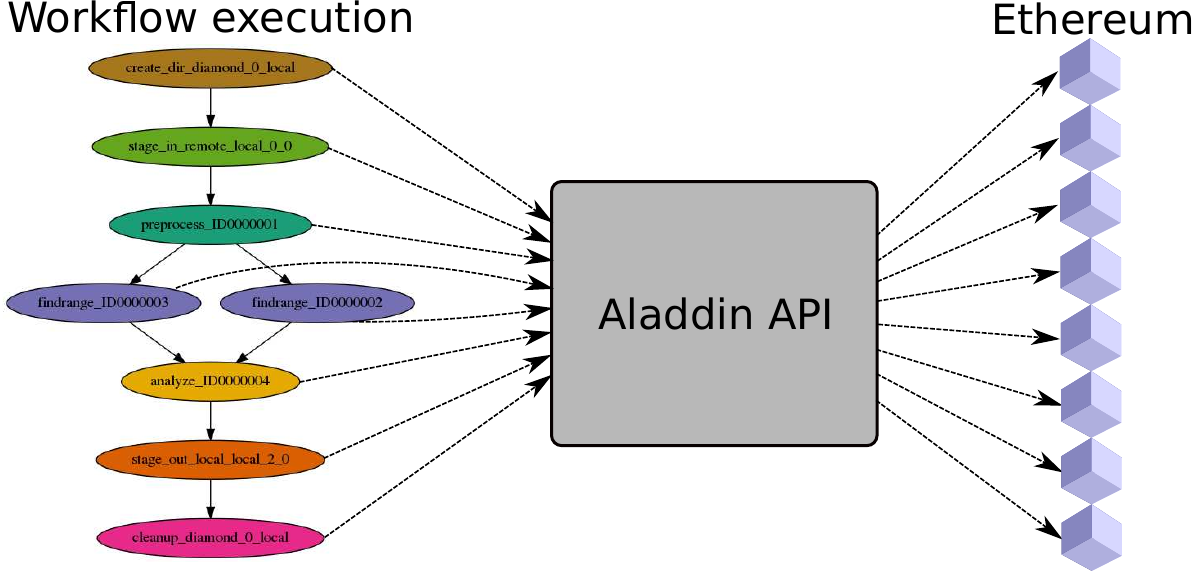}
    \caption{Using Aladdin API to log Pegasus workflow execution events to Ethereum blockchain.}
    \label{fig:workflow}
    \vspace{-2mm}
\end{figure}

When a workflow is submitted to Pegasus, it is decomposed in smaller tasks that are organized in a graph, so that their dependencies can be tracked and these tasks can be distributed across the execution environment. One of such sub-tasks is the staging of data into the computation nodes for execution (the second block in the workflow execution graph in Fig. \ref{fig:workflow}). Before transferring input files to the execution node, Pegasus generates a SHA-256 hash on each of them, to verify if the copy succeeded without errors. We added an extra step to this process, which is to register the file transfer along with the hashcodes into Ethereum, making that information instantly shared and non-repudiable. This way, other researchers can verify their hashcodes and easily detect errors in the data.

Like in the data staging, other execution tasks are also logged to the blockchain. The consolidation of logs has the potential to facilitate future analyses on the execution of workflows, for instance: find the more efficient execution environment or computing node, compare execution times, or detect possible anomalies in the execution environment. Information from the Pegasus log subsystem is encoded in JSON format and written to the blockchain once every sub-task is concluded.
We have published the formatted results of the blockchain logging at \url{https://bit.ly/2Mx3zO3}, where job details can be verified along with the information about the transaction in which they were recorded.

\vspace{-1mm}
\section{Identifying Anomalies in Blockchain Transactional Behavior using Machine Learning}
\label{sec:anomalies}

Initial coin offering (ICOs) and Security Token Offering (STOs) are recently changing the landscape of funding small businesses and startup ideas across the marketplace.  Both STOs and ICOs are built utilizing Smart Contracts (SC), which run on the blockchain and automatically execute when certain requirements are met.  
This differentiates such applications from traditional businesses because SCs identifiers can be tracked to expose their transactions. Consequently, such an approach can provide a clear picture of transactions surrounding a particular business and how and with whom it operates. 
Regulatory entities, such as the SEC, can take advantage of data analytics to better monitor ICOs and STOs, as they begin their operations.  Such transparency and introspection 
will not only facilitate easier regulation and monitoring of such companies, but it will also lead to more trusted investment opportunities for individuals as this space progresses. 

Given such needs that arise from this modern scenario of growing adoption of blockchain technologies, we pose the question whether it is possible to automate the identification of suspicious activity in blockchains. To this end, we propose the use of clustering techniques and SVMs to identifying outliers through the automated identification of discriminating features, that differentiate the outliers from the more conventional transactions between Ethereum accounts and smart contracts.

\vspace{-1mm}
\subsection{Experiments}

We define an outlying account as an account whose transaction activity is somehow different than the vast majority of the rest of the accounts. The largest group of accounts with similar transaction activity are considered to have normal transaction profiles.

Our approach compares different unsupervised machine learning algorithms to cluster averaged account transactions across a timeline into two categories: outliers and normal accounts. Using the results from the machine learning algorithms we then search the outliers using public websites that identify rogue accounts, Etherscan \footnote{\url{https://etherscan.io/}} and CryptoScamDB\footnote{\url{https://cryptoscamdb.org/}} (formerly EtherscamDB). We found that the approach did in fact discover rogue accounts, consistent to public comments on those accounts. 

Using publicly available transactions, account features were constructed aggregating transaction information and the age of each account. These account features were put into a dataset containing about 7 million accounts. These were then used to train machine learning algorithms for outlier detection. The first machine learning algorithm used for this was an SVM for novelty detection.

A One Class SVM, or OCSVM \cite{Scholkopf2001}, is an unsupervised machine learning algorithm that detects anomalies by learning a spatial boundary in which most of the samples are contained. If the sample is located outside of those boundaries, it is considered a novelty. Training the OCSVM on the account features produced accounts that were flagged as outliers. A number of these outliers had comments in Etherscan suggesting that they are scam accounts.

As a way of seconding the results of the OCSVM, another machine learning algorithm was used on the same data, K-Means clustering. K-Means clustering is another unsupervised algorithm that partitions the observations into a predefined (\emph{k}) number of clusters, and each observation is clustered with the nearest cluster mean. K-Means yielded a lower number of accounts considered to be outliers compared to the OCSVM, but 85\% of the outlying accounts produced from the K-Means were also flagged as outliers by the OCSVM. Public comments from Etherscan affirm that the suspected accounts are scams. With the data from the K-Means, another machine supervised learning algorithm could be implemented to check the consistency of both the OCSVM and K-Means.
 
\begin{table}[t]
 \caption{Results of unsupervised and supervised clustering using a test partition of the data}
 \begin{center}
 \begin{tabular}{|c|c|c|c|c|c|}
 \hline
 \multicolumn{6}{|c|}{Grouping Results} \\
 \hline
 \multicolumn{2}{|c|}{} & \multicolumn{4}{|c|}{K-Means} \\ \cline{3-6}
 \multicolumn{2}{|c|}{} & G1 & G2 & G3 & G4 \\
 \hline
 \multirow{4}{3em}{SVM} 
 & G1 & 5588166 & 3 & 0 & 9 \\ \cline{2-6}
 & G2 & 0 & 16 & 0 & 0 \\ \cline{2-6}
 & G3 & 1 & 0 & 4 & 2 \\ \cline{2-6}
 & G4 & 10 & 0 & 0 & 79 \\
 \hline
 \end{tabular}
 \label{tab:confusion}
 \end{center}
\vspace{-4mm}
\end{table}

In another experiment, the labeled clusters from the K-Means are then used as training labels for an SVM, used to classify the data. Similar to OCSVM, an SVM is a supervised machine learning algorithm that tries to find optimal separation margins in the data it is given, to separate it into classes. When the remaining test subset of the data is run through the SVM, the predicted classes match the groupings created by the K-Means by about 86\% as seen in the confusion matrix presented in Table \ref{tab:confusion}.

\begin{figure}[hpt]
\vspace{-4mm}
    \includegraphics[width=1\linewidth,trim=0 10 0 10,clip]{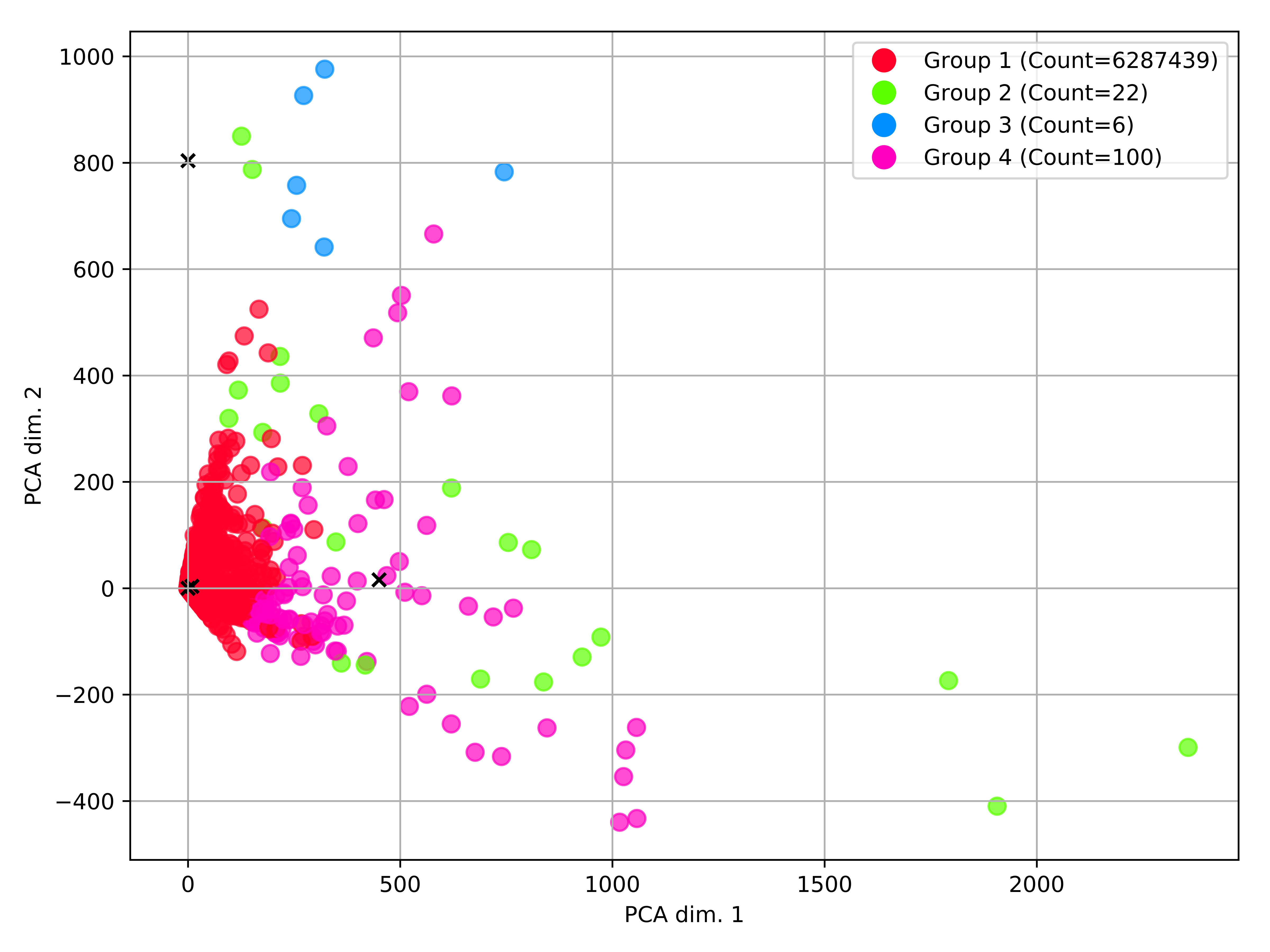}
    \caption{K-Means Cluster results of Ethereum account data. The black Xs denote cluster centers.}
    \label{fig:kmeanscluster}
\end{figure}

However different in their workings, all three algorithms had overlap with each other. These overlaps are the accounts flagged as outliers. The OCSVM flagged the highest amount of accounts as outliers.
The K-Means used four centers to group the ethereum account data (Seen as Xs in Fig. \ref{fig:kmeanscluster}), and OCSVM agreed regarding these outliers. The SVM classified the data based on the results from the K-Means where it yielded results similar to the K-Means.

\vspace{-6mm}
\section{Conclusions}
\label{sec:conclusions}
In this paper, we described three areas of work surrounding the the ingestion, retrieval and analysis of Blockchain data.  We described a system we developed that is capable of crawling the blockchain in order to extract data and to cross correlate the Smart Contract interfaces with the code in order to reconstruct the metadata correctly.  We discussed lessons learned from this work for public blockchains where scalability becomes quite challenging, but maintaining that such a solution could be used to retrieve data from private blockchains, using the elasticity of the cloud to meet such demands. We then described an implementation for collecting data about the execution and data used in scientific workflows, by augmenting the Pegasus workflow systems and harvesting data from tasks and dependencies in a generic way.  Finally, we discussed how we applied a number of machine learning techniques to identify abnormal activity in public Ethereum accounts. Currently, we can process the entire Ethereum dataset, and we showed how we successfully identified outlier accounts that warrant further investigation.   We are also collecting and organizing external information that will allow us to perform more reliable evaluation of these methods.

\bibliographystyle{IEEEtran}
\bibliography{IEEEabrv,references}

\end{document}